\documentclass[twocolumn,tighten]{aastex63}
\RequirePackage{rotating}
\usepackage{xfrac}

\newcommand{\kms}{${\rm km\,s^{-1}}$}

\newcommand{\cc}{$^{13}$CO}
\newcommand{\ce}{C$^{18}$O}

\shorttitle{Radial variation of CO gas abundance}
\shortauthors{Zhang et al.}

\begin{document}

\title{Systematic Variations of CO Gas Abundance with Radius in Gas-rich Protoplanetary Disks}

\correspondingauthor{Ke Zhang}
\email{kezhang@umich.edu}

\author[0000-0002-0661-7517]{Ke Zhang}
\altaffiliation{Hubble Fellow}
\affil{Department of Astronomy, University of Michigan, 323 West Hall, 1085 S. University Ave, Ann Arbor, MI 48109, USA}

\author[0000-0003-4179-6394]{Edwin A. Bergin}
\affil{Department of Astronomy, University of Michigan, 323 West Hall, 1085 S. University Ave, Ann Arbor, MI 48109, USA}

\author[0000-0002-6429-9457]{Kamber R. Schwarz}
\altaffiliation{Sagan Fellow}
\affil{Lunar and Planetary Laboratory, University of Arizona, 1629 E. University Blvd, Tucson, AZ 85721, USA}

\author[0000-0002-3291-6887]{Sebastiaan Krijt}
\altaffiliation{Hubble Fellow}
\affil{Department of Astronomy/Steward Observatory, The University of Arizona, 933 North Cherry Avenue, Tucson, AZ 85721, USA}

\author[0000-0002-0093-065X]{Fred Ciesla}
\affil{Department of the Geophysical Sciences, The University of Chicago, 5734 South Ellis Avenue, Chicago, IL 60637, USA}

\begin{abstract}

CO is the most widely used gas tracer of protoplanetary disks. Its abundance is usually assumed to be an interstellar ratio throughout the warm molecular layer of the disk. But recent observations of low CO gas abundance in many protoplanetary disks challenge our understanding of physical and chemical evolutions in disks. Here we investigate the CO abundance structures in four well-studied disks and compare their structures with predictions of chemical processing of CO and transport of CO ice-coated dust grains in disks.  We use spatially resolved CO isotopologue line observations and detailed thermo-chemical models to derive CO abundance structures. We find that the CO abundance varies with radius by an order of magnitude in these disks. We show that although chemical processes can efficiently reduce the total column of CO gas within 1\,Myr under an ISM level of cosmic-ray ionization rate, the depletion mostly occurs at the deep region of a disk. Without sufficient vertical mixing, the surface layer is not depleted enough to reproduce weak CO emissions observed. The radial profiles of CO depletion in three disks are qualitatively consistent with predictions of pebble formation, settling, and drifting in disks. But the dust evolution alone cannot fully explain the high depletion observed in some disks. These results suggest that dust evolution may play a significant role in transporting volatile materials and a coupled chemical-dynamical study is necessary to understand what raw materials are available for planet formation at different distances from the central star. 
\end{abstract}


\section{Introduction}
\label{sec:introduction}

Planets are formed of raw materials in protoplanetary disks around young stars \citep{Hayashi81, Williams11}.  The mass distribution and composition of solids and gases in natal disks profoundly affect outcomes of planetary systems, from planetary masses to compositions of planetary cores and atmospheres \citep{Pollack96, Oberg11, Benz14}. Recently, high spatial resolution observations of dust continuum and gas lines in disks start providing direct constraints to on-going planet formation processes \citep{alma15, Andrews16, Andrews18b}. To accurately constrain initial conditions and intermediate processes of planet formation, we need a robust understanding of the physical and chemical processes relevant to observational tracers \citep{Bergin18_mass}. 

Among all gaseous tracers of protoplanetary disks, CO is the most widely detected and studied molecule.  It has the advantages of being highly abundant, relatively chemically stable, and readily detectable \citep{Molyarova17}. The typical view of CO gas distribution in protoplanetary disks is based on two concepts: \textit{ the snowline} and \textit{ the warm molecular layer} \citep{Aikawa02}. Along the radial direction, CO exists as the gas inside its snowline and freezes out as ice beyond the snowline. In the vertical direction, the CO gas presents in a middle layer, the so-called warm molecular layer, of which the lower boundary is where CO freezes out, and the upper boundary is where CO is photo-dissociated by stellar or external radiation. Inside the warm molecular layer, the CO gas abundance is assumed to be a constant across all radii, typically at an interstellar ratio of $\sim10^{-4}$. This simple parameterization has been widely used in constraining basic properties of protoplanetary disks, such as the total disk mass, \citep{Williams14, Ansdell16}, gas disk radius \citep{Andrews12, Ansdell18}, gas mass distribution \citep{Williams16, Zhang17, Miotello18}, gas temperature \citep{Salyk09, Flaherty15, Schwarz16}, and kinematic properties due to turbulence and planet-disk interactions \citep{Flaherty15, Teague18a, Pinte18b}.

 \begin{figure*}[!htbp]
\centering
\includegraphics[width=0.8\textwidth]{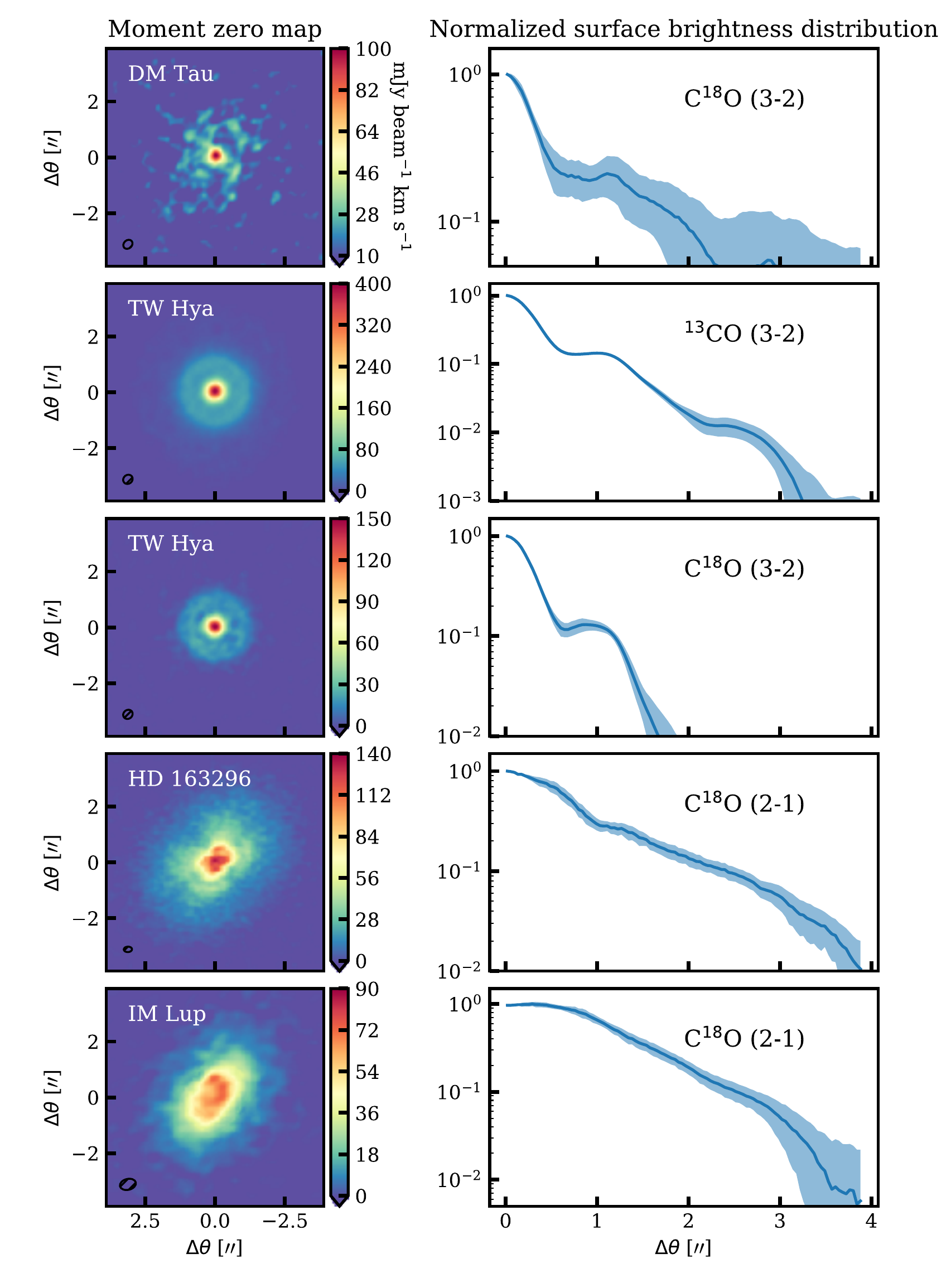}
\caption{Left column: Moment zero map (integrated intensity) of CO isotopologue line observations (continuum subtracted). The axes are labeled with angular offsets from the disk center, and the synthesized beam is shown in the lower-left corner of each panel. RIght column:  deprojected radial line intensity profiles of each line.  The shaded areas indicate a 1$\sigma$ uncertainty. The deprojections used the disk geometric parameters listed in Table~\ref{tab:stellar_paras}.  \label{fig:observations}}
\end{figure*}

However, this simple picture of CO in disks has been challenged by growing numbers of observations. Systematic (sub)mm surveys of Class II disks in nearby star formation regions showed that the CO gas-to-dust mass ratios in the majority of  Class II disks are one order of magnitude too low compared to the interstellar ratio, even after correcting the effects of CO freeze-out and isotope-selective photodissociation \citep{Ansdell16, Barenfeld16, Miotello16, Long17}.

It is still under debate whether the weak CO lines are due to gas dissipation or a sub-interstellar CO abundance, but the latter case has received circumstantial support.  All three disks with independent gas mass measurements from HD line fluxes show a significant depletion in CO abundance (by a factor of 5-100) while their hydrogen gas is still abundant \citep{Bergin13, Favre13, McClure16, Schwarz16, Zhang17}. Also, the measured mass accretion rates and expected ages imply an ISM level gas-to-dust ratio of 100 for disks in the Lupus star formation region \citep{Manara16}. Furthermore, no correlation was found between the CO derived gas masses and M$_\star$, in contrast to a clear correlation between dust mass and M$_\star$ \citep{Miotello17}. These results suggest that CO depletion, if not the only reason for the weak CO emission, may occur widely in disks. Its cause is likely a fundamental process in disk evolution and planet formation. 

Various chemical and physical processes have been proposed to explain CO depletion observed in protoplanetary disks. Chemical processes can reprocess CO into other molecules and gradually reduce the CO abundance in disks \citep{Aikawa99, Bergin14, Reboussin15, Eistrup16, Yu16, Schwarz18, Bosman18, Dodson18}. On the other hand, warm CO gas diffuses into colder region closer to the mid-plane, freezes on dust grains, and is consequently sequestered in the mid-plane as dust grains grow and settle  \citep{Krijt16, Xu17}.  As grains radially drift inward, they further redistribute condensed volatiles in the disk \citep{Oberg16, Stammler17, Krijt18}. Even starting with a constant CO abundance in the disk,  these chemical and physical processes will lead to a varying CO gas abundance in the mid-plane and the warm molecular layer of the disk \citep{Krijt18}. 

However, previous observational studies mostly focused on disk-averaged CO abundance, and very few studies compared the predictions of depletion mechanisms directly with spatially resolved observations. 
Here we make the first attempt to investigate whether the CO depletion varies with radius. Our approach is to compare spatially resolved CO isotopologue line observations with detailed thermo-chemical models for four well-studied disks.  We mainly use \ce~line observations, as $^{12}$CO and $^{13}$CO lines are expected to be highly optically thick in most regions of the disk and \ce~low $J$ lines are mostly optically thin beyond $\sim$30\,AU.

This paper is structured as follows: in Section~\ref{sec:obs}, we present the observations of line images and describe the data calibration processes.  Section\,\ref{sec:methods} explains the detailed chemical modeling and choices of input parameters. In Section~\ref{sec:results}, we compare the chemical models with observations and derive CO gas abundance distributions needed to match with observations. In  Section~\ref{sec:discussion}, we discuss how the current models of chemical processing and dust evolution in disks may produce the radial CO abundance profiles derived here. Then we summarize the findings in Section~\ref{sec:summary}.

\section{Observations}
\label{sec:obs}
We employed a sample of 4 protoplanetary disks with high angular resolution \ce~line observations from ALMA archive. The four disks are DM Tau, TW Hya, HD 163296, and IM Lup. These sources were selected based on availability and quality of their \ce~line observations. The spatial resolution of these observations is between 0.\arcsec2 and 0.\arcsec5, corresponding to 20-80\,AU resolution based on distances from Gaia measurements \citep{Gaia1, Gaia2}.  For the TW Hya disk, we included additional \cc~(3-2) line observations to constrain the CO abundance structure in $>$100\,au region, where its \ce~(3-2) emission drops below the noise level.  The usage of \cc~line for TW Hya is based on previous studies which showed that the \cc~(3-2) line of the TW Hya disk is optically thin beyond 70\,au \citep{Schwarz16}. We did not include \cc~line observations for the other three disks, because previous studies suggested that \cc~low $J$ lines of these disks are likely to be highly optically thick at most of the disk regions \citep{Williams14, Flaherty15, Cleeves16}. A journal of the ALMA observations is provided in Table~\ref{tab:obslog}.

All raw data were calibrated through the accompanying reduction scripts under the corresponding version of CASA \citep{McMullin07}. Bandpass and phase calibration used bright quasars as calibrators, and the flux calibrations were based on either a Solar System object or bright quasars. The absolute flux uncertainty is estimated to be 10\%. 

The following imaging processes were all performed with CASA 5.0. We carried out self-cal for each data set using continuum visibilities combined from line-free channels. For IM Lup, two datasets from different programs were combined together after individual self-cal, using  \texttt{fixvis} and \texttt{fixplanets} tasks in CASA to shift the data sets to a common phase center. The calibrated data were imaged in CASA using the \texttt{tclean} task. Briggs weighting (robust=0.5 or 1) and the multiscale option with a factor of (0,\,5,\,10) were used. The channel widths used for images are 0.2 or 0.22\,\kms~(see Table~\ref{tab:f_line}), similar to the spectral resolution in the original observations, except for HD 163296. Although the original observations of HD 163296 have a higher spectral resolution, we binned the data into a channel width of 0.2\,\kms for a better signal-to-noise ratio. The noise levels and line fluxes of the channel maps can be seen in Table~\ref{tab:f_line}.

\begin{deluxetable*}{ccrccccc}[htpb!]
\tablecaption{Observation log \label{tab:obslog}}
\tablecolumns{8}
\tablewidth{0pt}
\tablehead{
\colhead{Source} & \colhead{Project ID} & \colhead{PI} &
  \colhead{t$_{\rm int}$} & 
\colhead{Baseline} & \colhead{Antenna} &\colhead{UT time} &
\colhead{Channel width}  \\
 &  & &
  \colhead{(min)} &
   \colhead{(m)} & &
 & \colhead{(kHz)} 
}
\startdata
DM Tau	&	2015.1.00308.S	&	E. A. Bergin	&		41	&	15-1091	&	40-54	&	07/2016	&121		\\														
TW Hya	&	2016.1.01495.S	&	H. Nomura	&		52	&	15-704	&	45	&	12/2016	&	141		\\
HD 163296	&	2013.1.00601.S	&	A. Isella	&		151	&	35-1574	&	44	&	08/2015	&	31	\\														
IM Lup	&	2013.1.00226.S 	&	K. Oberg	&		21	&	20-650	&	31	&	07/2015	&	61		\\														
	       &	2013.1.00798.S 	&	C. Pinte	&		37	&	21-784	&	36	&	06/2015	&	35	\\
\enddata

\end{deluxetable*}

\begin{deluxetable*}{ccccccc}[htbp!]
\tablecaption{Line information \label{tab:f_line}}
\tablecolumns{7}
\tablewidth{0pt}
\tablehead{
\colhead{Source} & \colhead{Line} & \colhead{$\Delta v$} &
\colhead{rms} & 
\colhead{F$_{\rm int}$} & \colhead{Beam}  & \colhead{Beam} \\
 &  &  \colhead{(\kms)} & \colhead{(mJy/beam)} &\colhead{(Jy \kms)} &  \colhead{(arcsec, PA)} &  \colhead{(AU)}
}
\startdata
DM Tau	&	\ce\,(3-2)	&	0.22	&			5	&	3.0$\pm$0.1		&	0.36$\times$0.29 (-47) & 50$\times$41\\
TW Hya	&	\ce\,(3-2)	&	0.22	&			5.3	&	1.21$\pm$0.04		&	0.36$\times$0.31 (-51) &22$\times$19\\
	       &	\cc\,(3-2)	&	0.22	&			4	&	4.61$\pm$0.05		&	0.36$\times$0.32 (-52)&22$\times$19\\
HD 163296	&	\ce\,(2-1)	&	0.2	&			2.86	&	7.3$\pm$0.1		&	0.29$\times$0.20 (-88)& 29$\times$20\\
IM Lup	&	\ce\,(2-1)	&	0.2	&			4	&	1.29$\pm$0.04		&	0.58$\times$0.39 (-76)& 93$\times$62\\
\enddata
\end{deluxetable*}

\begin{deluxetable*}{cccccccccccc}[htbp!]
\tablecaption{Source information \label{tab:stellar_paras}}
\tablecolumns{12}
\tablewidth{0pt}
\tablehead{
\colhead{Source}  &  \colhead{M$_\star$}  &  \colhead{R$_\star$}  &  \colhead{T$_{\rm eff}$}  &  \colhead{D}  &  \colhead{M$_{\rm d}$ }  &  \colhead{M$_{\rm gas}$}  &  \colhead{V$_{\rm lsr}$}  &  \colhead{incl}  &  \colhead{PA}  &  \colhead{L$_X$}  &  \colhead{Refrences}\\
\colhead{}  &  \colhead{(M$_\odot$)}  &  \colhead{(R$_\odot$)}  &  \colhead{(K)}  &  \colhead{(pc)}  &  \colhead{(M$_\odot$)}  &  \colhead{(M$_\odot$)}  &  \colhead{(\kms)}  &  \colhead{(deg)}  &  \colhead{(deg)}  &  \colhead{(erg s$^{-1}$ cm$^{-2}$)}  &  \colhead{}
}
\startdata
DM Tau  &  0.53  &  1.25  &  3705  &  140  &  5.00E-04  &  5.00E-02  &  5.95  &  35  &  335  &  3.00E+29  &   1,\,2,\,3,\,4,\,5\\
TW Hya  &  0.8  &  1.04  &  4110  &  60  &  5.00E-04  &  5.00E-02  &  2.84  &  7  &  335  &  1.60E+30  &  2,\,6,\,7,\,8  \\
HD 163296  &  2.47  &  1.69  &  9250  &  101  &  1.40E-03  &  1.50E-01  &  5.74  &  49  &  312  &  4.00E+29  & 2,\,9,\,10,\,11,\,12,\,16  \\
IM Lup  &  1  &  2.5  &  3900  &  160  &  1.70E-03  &  1.70E-01  &  4.5  &  48  &  144  &  4.30E+30  &   2,\,13,\,14,\,15\\
\enddata

\tablecomments{\scriptsize (1)\,\citet{Kenyon95}, (2)\,\citet{Gaia2}, (3)\,\citet{Andrews11}, (4)\,\citet{Bergin16}, (5)\,\citet{Henning10},
(6)\,\citet{Andrews12}, (7)\,\citet{Huang18a}, (8)\,\citet{Brickhouse10}, 
(9)\,\citet{Natta04}, (10)\,\citet{Isella16}, (11)\,\citet{Rosenfeld13b}, (12)\,\citet{Swartz05},
(13)\,\citet{Pinte08}, (14)\,\citet{Cleeves16}, (15)\,\citet{Gunther10}, (16)\,\citet{deGregorio-Monsalvo13}.
}
\end{deluxetable*}

\begin{deluxetable*}{lcccccl}
\tablecaption{Disk Model Parameters \label{tab:disk_paras}}
\tablecolumns{7}
\tablewidth{0pt}
\tablehead{
\colhead{}  &  \colhead{DM Tau}  &  \colhead{TW Hya}  &  \colhead{HD 163296}  &  \colhead{IM Lup}   & \colhead{Units}&  \colhead{Definition}}
\startdata
$\gamma_{\rm g}$  &  1  &  0.75  &  0.8  &  1  &  &surface density index of gas \\
$\gamma_{\rm mm}$  &  1  &  1  &  0.1  &  0.3  &  &surface density index of large grain \\
$R_c^{g}$  &  270  &  104  &  165  &  100  &  (AU) &characteristic radius of gas \\
$R_c^{\rm mm}$  &  135  &  60  &  90  &  100  &   (AU) &characteristic radius of large grain \\
$M_{\rm mm}$  &  4.00E-04  &  4.00E-04  &  1.20E-03  &  1.70E-03  &(M$_\odot$)  &mass of large grain \\
$M_{\rm small}$  &  1.00E-04  &  1.00E-04  &  3.00E-04 &  1.70E-05  & (M$_\odot$) &mass of small grain \\
$\chi_{\rm mm}$  &  0.2  &  0.2  &  0.2  &  0.2  &  &scale fraction of large grain \\
H$_{100}$ (AU)  &  6  &  6  &  8.5  &  8  &  (AU) &gas scale height at 100AU \\
$\psi$  &  1.25  &  1.25  &  1.08  &  1.2  &  &flaring parameter \\
R$_{in}^{\rm mm}$  &  19  &  1  &  0.5  &  0.2  & (AU)  &inner radius of large grain \\
R$_{in}^{\rm g}$  &  3  &  4  &  0.5  &  0.2  &  (AU) &inner radius of gas/small grain \\
R$_{out}^{\rm mm}$   &  300  &  70  &  600  &  313  & (AU)  &outer radius of large grain \\
R$_{out}^{\rm g}$  &  500  &  200  &  600  &  1200  &  (AU) &outer radius of gas/small grain \\
\enddata

\end{deluxetable*}

\section{Methods}
\label{sec:methods}

The goal of this study is to investigate how the CO gas abundance varies with radius in the warm molecular layer of a protoplanetary disk. Our approach is to analyze a sample of well-studied disks in a homogeneous framework, by comparing spatially resolved emission of optically-thin CO isotopologue lines to the predictions of thermo-chemical models of constant carbon elemental abundance across disks.

In the following sections, we describe our parameterization on the density structures in protoplanetary disks, assumptions on the input parameters and previous constraints, the numerical methods for radiative transfer and chemical-evolution, and our approach for comparing models with observations.

\subsection{Parametric Density Model}

We employ a global surface density from the self-similar solution of a viscously evolving disk \citep{Lynden-Bell74}, see eq.~(\ref{eq:sigma_profile}).  This profile has been widely used in modeling of protoplanetary disks \citep[e.g.][]{Andrews11, Zhang14, Cleeves16}. 

\begin{equation}\label{eq:sigma_profile}
\Sigma (R) = \Sigma_c \Big(\frac{R}{R_c}\Big)^{-\gamma} {\rm exp} \Big[ -\Big(\frac{R}{R_c}\Big)^{2-\gamma}\Big] 
\end{equation}

where $\Sigma_c$ is the surface density at the characteristic radius R$_c$, and $\gamma$ is the gas surface density exponent. 

The disk model is composed of three mass components -- gas, a small dust population, and a large dust grain population. The separation of different dust populations is used to mimic the effects of dust evolution in protoplanetary disks, including vertical settling and radial drifting \citep{birnstiel12,Krijt16}. The mass distribution of all three components are assumed to follow the global surface density profile as eq.~(\ref{eq:sigma_profile}), but each component can have a different set of parameter values. The gas and the small grain population are assumed to be spatially coupled, while the large grain population has a different spatial distribution.  

For each mass component, the vertical distribution is characterized by a scale height $h(R)$ and the density $\rho$ is computed through eq.~(\ref{eq:rho}). 
\begin{equation}\label{eq:rho}
\rho_i(R, Z) = f_i \frac{\Sigma (R)}{\sqrt{2\pi} h_i(R)}  {\rm exp} \Big[ -\Big(\frac{Z}{h_i(R)}\Big)^2\Big], 
\end{equation}

\begin{equation}\label{eq:hr}
h_i (R) = \chi_i h_0 (R/100 {\rm AU} )^\psi 
\end{equation}
where $f_i$ is the mass fraction of each mass component, h$_0$ is the scale height at 100\,AU, $\psi$ is a parameter that characterizes the radial dependence of the scale height. The large grain population is expected to be more settled compared to the gas and small grains. This is modeled by the settling parameter $\chi_i$, where $\chi$=1 of the gas and the small grain population, and $\chi <1$ for the large grain population. 

Both dust populations are assumed to follow an MRN size distribution $n(a)\propto a^{-3.5}$ with a minimum grain size a$_{\rm min}$ =0.005\,$\mu m$, and  a$_{\rm max}$ =1\,$\mu m$ for the small grain population and a$_{\rm max}$=1\,mm for the large grain population. We use Mie theory to compute the wavelength dependence of dust opacity.

\subsection{The adoption of input parameters and previous observational constraints}

\begin{figure*}[htbp]
\centering
\includegraphics[width=0.8\textwidth]{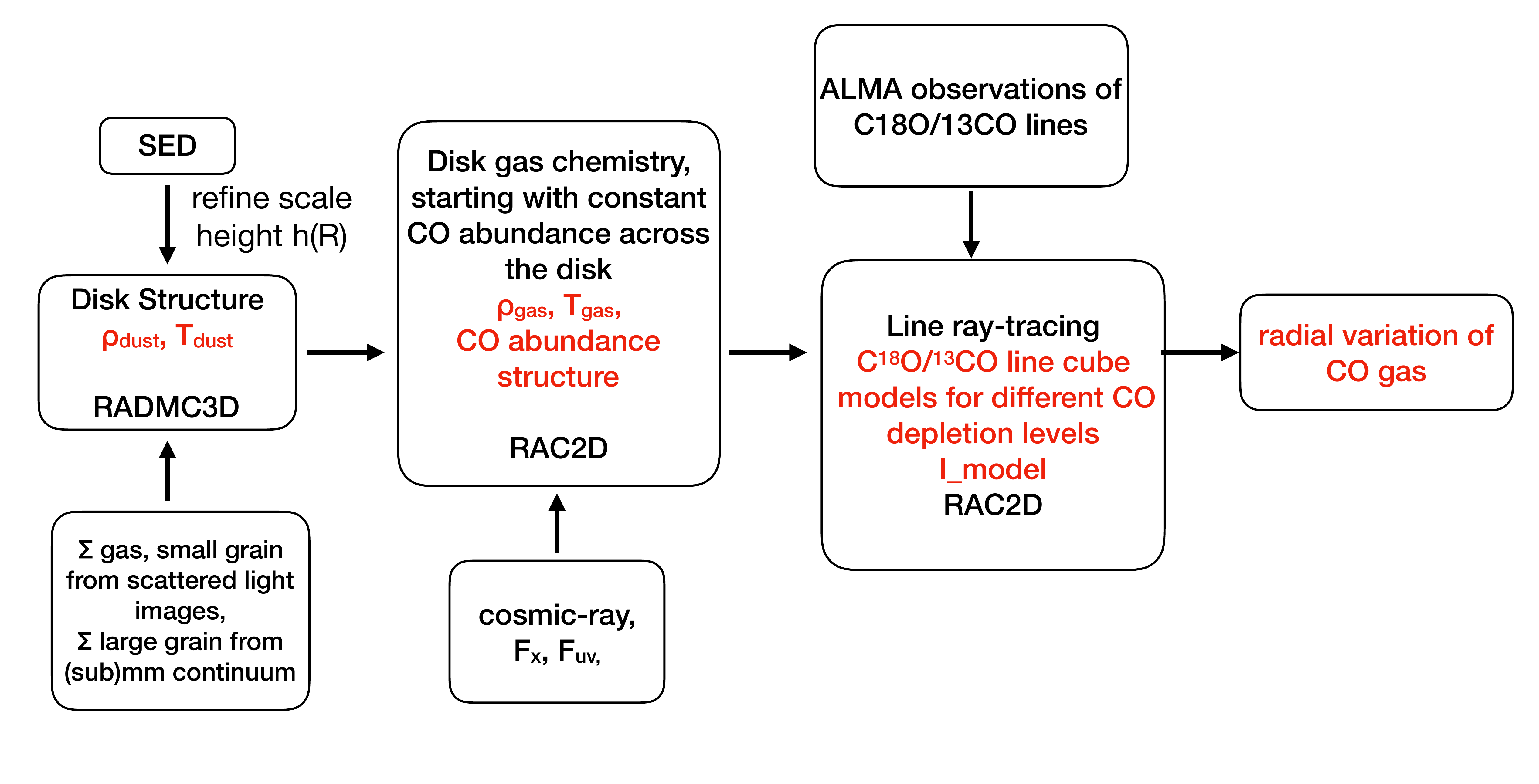}
\vspace{-0.1cm}
\caption{An outline of the modeling processes. The output of each step is highlighted in red. \label{fig:flowchart}}
\end{figure*}

\subsubsection{Disk parameters}
All four disks are well-studied disks and their disk properties have been under intensive studies in the literature.  For our purpose of studying CO abundance structure,  the most important parameters are the gas surface density profile and the disk flaring level. The former sets the baseline of CO gas distribution, and the latter largely determines the temperature structure in a disk. When gas/small dust distribution constraints are available from scattered light images, we adopt these values from previous studies. If not, we adopt $\gamma_{\rm g}$=1 for gas and small dust grains. The gas disk sizes are adopted from scattered light images or previous studies of multiple CO isotopologue lines.  We vary the $\gamma$ index of gas surface density profiles to test the uncertainties in the derived CO abundance structures in Section~\ref{sec:results}.  

For a given surface density profile, the flaring level of the disk can be constrained by fitting its broad-band spectral energy distribution \citep[SED,][]{dullemond04}. 
 The mass ratio of the small-to-large grains and the settling of large grains can affect the thermal structure and the UV penetration in a disk.  \citet{Schwarz18,Schwarz19a} had carried out systematic tests on the effect of small-to-large dust mass ratio on the CO processing chemistry. They showed that the CO depletion happened faster in models with higher small-to-large dust mass ratio. Here we adopt an intermediate level of 80\% of the mass in large grains for modeling. The only exception is the IM Lup disk for which we adopt 99\% of the mass in large grains as this value was used on the comprehensive model of the IM Lup disk of \citet{Cleeves16}. Here we use a settling parameter of $\chi$=0.2 for the large grain population.  The detailed disk structures inside 20\,AU is not critical for our analysis as we focus on CO gas distribution on a scale of tens of AU. We also ignore small-scale substructures in these disks \citep{Andrews16, Zhang16, Andrews18b},  as this work focuses on the large-scale radial variations.  Please see section~\ref{dis:substructure_effects} for a more detailed discussion on possible effects of substructures on CO abundance distribution.
 
 Two of these four disks (TW Hya and DM Tau) have HD (1-0) line observations and previous models suggest that their gas-to-dust mass ratios are close to 100 \citep{Bergin13, McClure16}. For simplicity, we use a gas-to-dust (small+large) mass ratio of 100 for all four disks. The uncertainty in the total gas mass is not crucial for the study here, as we focus on the radial variation patterns. 
  
Table~\ref{tab:disk_paras} lists all the parameters adopted for our disk models. Here we briefly describe the adoption of parameters for each disk: 
\begin{itemize}
   \item DM Tau: we employ a $\gamma_{\rm g}$=1 for gas and small grains, as no scattered light image studies are available. The inner edge of gas is set to be 4\,AU \citep{Calvet05}. The characteristic radius of gas, R$_c^{\rm g}$, is set to 270\,AU based on the extent of \ce\,(3-2) emission in channels close to the stellar velocity.  For the large grain population, we use $\gamma_{\rm mm}$=1, R$_c^{\rm mm}$=135\,AU, and an inner radius of 20\,AU based on the model of \citet{Andrews11}, which matches the  880$\mu m$ continuum observations of DM Tau.
  \item TW Hya: we adopt the general gas surface density profile of TW Hya derived from scattered light images from \citet{Vanboekel17}. This distribution has a global $\gamma_{\rm g}$=0.7 and a exponential taper beyond 104 AU. The distribution of mm-grain is truncated at 70\,AU, with a R$_c^{\rm mm}$=60\,AU \citep{Andrews12}. 
  \item HD 163296: we employ the basic structure of \citet{Isella16} which fitted the 1.3\,mm continuum emission and the general sizes of CO/\cc/\ce\,(2-1) isotopologue emissions. The model uses R$_c^{\rm g}$ = 165\,AU and $\gamma_{\rm g}$=0.8 for gas and small grains,  and R$_c^{\rm mm}$ = 90\,AU, $\gamma_{\rm mm}$=0.1 for the large grain population. 
  \item IM Lup: we adopt  $\gamma_{\rm g}$=1 for gas, and other surface density profile parameters from \citet{Cleeves16}, which carried out a comprehensive model of 875\,$\mu m$ continuum and multiple CO\,(2-1) isotopologue emissions.
\end{itemize}

\subsubsection{Stellar parameters}
We adopt the effective temperatures and luminosities of the central stars from the literature and rescale the stellar luminosities using the latest Gaia 2 measurements. The inclination and position angles are adopted from previous analysis of spatially resolved line and continuum observations. The UV and X-ray luminosities are from the literature. All stellar information and disk geometric parameters are listed in Table~\ref{tab:stellar_paras}. 

\begin{figure}[!t]
\centering
\vspace{.2cm}
\includegraphics[width=0.5\textwidth]{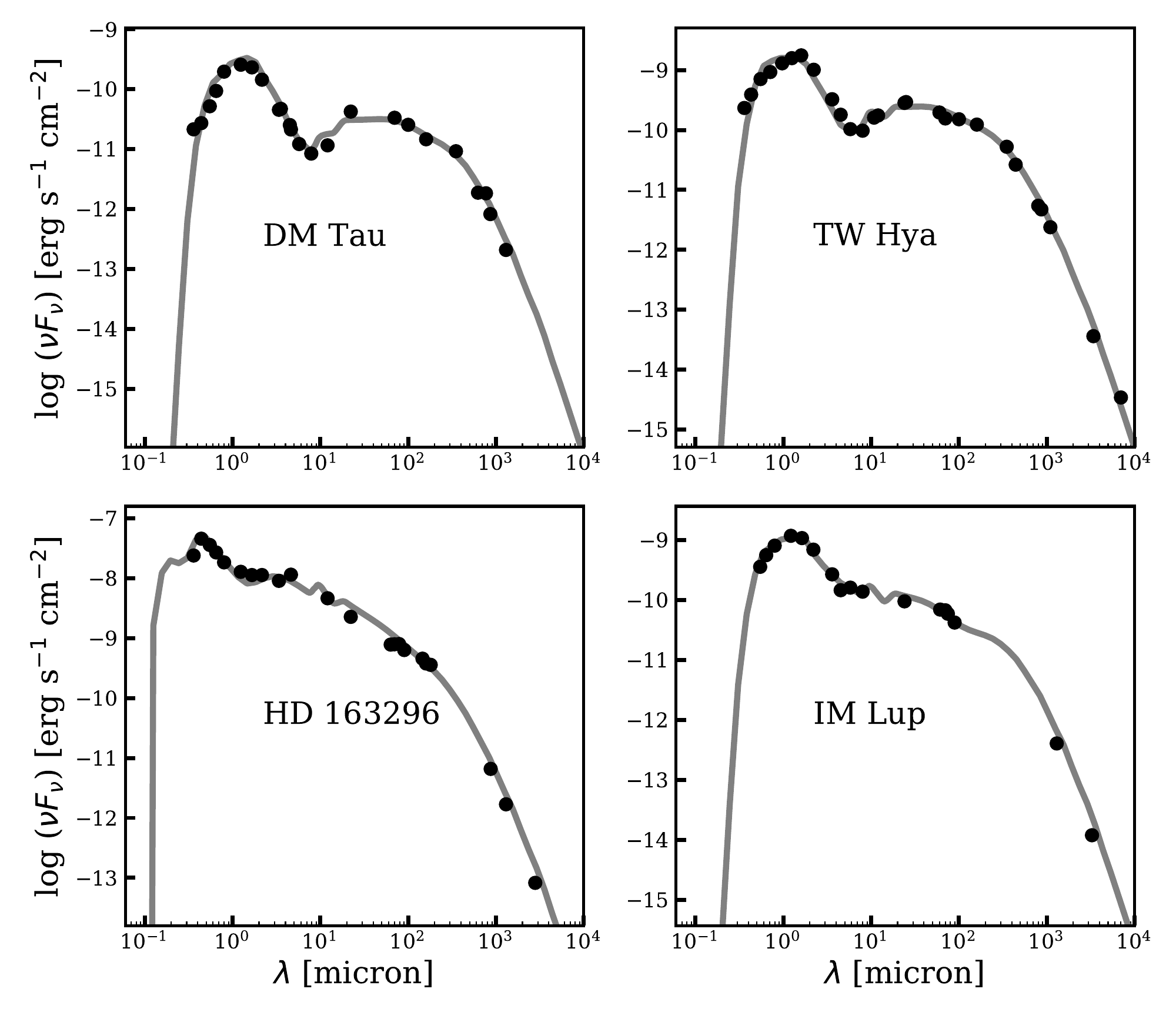}
\vspace{-0.3cm}
\caption{The best-fitting SED models of the four disks. The photometry data are black dots and the best-fittings models are overlaid as solid grey lines. \label{fig:seds}}
\end{figure}

\subsection{The numerical methods}

The general modeling processes are outlined in Figure~\ref{fig:flowchart}. 

We first use the radiative transfer code RADMC3D to calculate the dust thermal structure in a disk \citep{radmc3d} and then generate model SEDs to compare with observations. To constrain the vertical scale height of dust, we first run a grid of $h_0$ and $\psi$ to search for best-fitting SED models of each disk. The $h_0$ range is from 0.02 to 0.2 with a step size of 0.02, and $\phi$ from 1 to 1.3 with a step size of 0.05. 

Then we adopt the best-fitting values of density structure into a full 2D time-dependent thermo-chemical code RAC2D. This code computes the disk chemistry and the heating-cooling balance for both the gas and the dust self-consistently. We briefly describe the code here and interested readers can find a full description in \citet{Du14}.  Given a static density structure of gas and dust, the code first solves the dust temperature and radiation field (from X-ray to centimeter wavelengths) in the disk using a Monte Carlo approach. The cosmic ray ionization inside the disk is simulated with an attenuation length of 96 g cm$^{-2}$ \citep{Umebayashi81, Bergin07}.  For the given radiation field, the code then simultaneously solves the chemical evolution and gas thermal structure in the disk. The chemical network has 467 species and 4801 reactions, including the full gas-phase network from the UMIST database, dissociation of H$_2$O and OH by Ly$\alpha$ photons, adsorption and desorption of species on the dust grain surface either thermally or induced by cosmic rays and UV photons, and two-body reactions on the dust grain surface.

The chemical models are initialized with a composition listed in the Table~1 of \citet{Du14}. In particular, all carbon is initially in CO gas with an abundance of 1.4$\times$10$^{-4}$ relative to H atoms and a binding energy of 855\,K for CO ice \citep{Oberg05}. We let the chemistry run for 1\,Myr for all models. 
 No isotopologue fractionation is considered in the chemical network, as the fractionation is expected to be insignificant in massive disks like this sample \citep{miotello14}. We rescale the output CO abundance by the local ISM ratios of CO/\cc= 69, and CO/\ce=570 to generate CO isotopologue abundances \citep{wilson99}. 

Using the gas temperature and CO abundance structures, we compute line images using the ray-tracing module of RAC2D, assuming an LTE condition. We initially simulate the images at a 2\,AU spatial resolution and 10$\times$spectral resolution as the observations, and then convolve model images to the same spatial and spectral resolution as observations. 

\subsection{Parameter Tests}
We test model uncertainties of two important parameters: the cosmic-ray ionization rate ($\zeta_{\rm CR}$) and the $\gamma$ index for surface density profiles. 

Previous chemical studies suggested that an ISM level of cosmic-ray (CR) ionization rate ($\zeta_{\rm CR}\ge1.36\times10^{-17}$ s$^{-1}$) is critical for sufficient CO processing to other molecules \citep{Schwarz18, Bosman18}. This level of cosmic ray rate is typical in ISM but it is unclear whether this is also present in disks, as disk wind or magnetic fields in actively accreting T Tauri disk systems may attenuate the cosmic ray rate in disks \citep{Cleeves13a, Cleeves15}.  To test the impact of CR rate on our chemical models, we run two sets of models: a set of standard models with a $\zeta_{\rm CR}=1.36\times10^{-18}$ s$^{-1}$ and the other set with a high CR rate of 1.36$\times10^{-17}$ s$^{-1}$. 

We also test how uncertainties on the slope of surface density profiles would impact the constraints on CO depletion profile. Therefore, for each disk, we run two additional models with $\gamma$ varies by $\pm$0.2. The characteristic radii and total gas masses are kept the same.

\section{Results}
\label{sec:results}

\subsection{CO abundance and gas temperature structures from thermo-chemical models}

\begin{figure*}[!htbp] 
\centering
\includegraphics[width=1\textwidth]{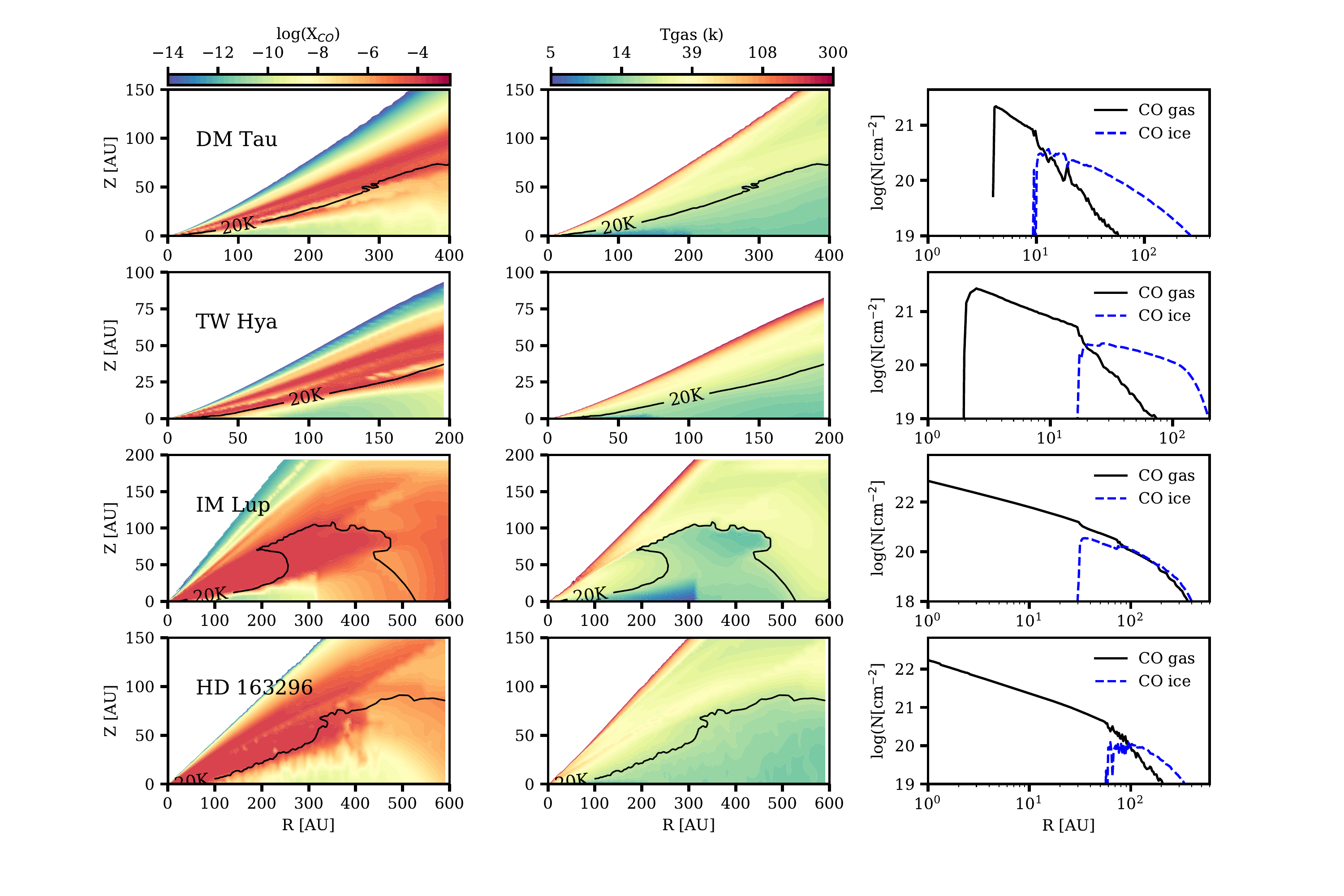}
\vspace{-0.9cm}
\caption{Chemical structures of the standard models ($\zeta_{\rm CR}$=1.36$\times$10$^{-18}$ s$^{-1}$). First column: CO gas abundance;  second column: gas temperature. The 20K contours are plotted to guide eyes for a general expectation of CO freeze-out location, but the actually freeze-out level depends on the local number densities of CO gas and ice as well as the temperature differences between gas and dust \citep{Du14}. The clear deviation from 20\,K is seen in the outer disk region ($>$300\,au) of IM Lup where the densities are low. The third column: column density profiles of CO gas and CO ice. \label{fig:standard_models}
 }
\end{figure*}

In Figure~\ref{fig:standard_models}, we show the CO gas abundance and gas temperature structures in the standard models ($\zeta_{\rm CR}=1.36\times10^{-18}$ s$^{-1}$) of these four disks. All four disks show a molecular layer in which the CO gas abundance is close to 10$^{-4}$.  This layered structure has been seen in channel maps of CO line observations of HD 163296 and IM Lup \citep{Rosenfeld13, Pinte18a}.  The last column in Figure~\ref{fig:standard_models} shows the column density distributions CO gas and ice in the four disk models.  The CO (gas+ice) column density distribution is close to 10$^{-4}$ at all radii, indicating very little CO has been converted to other carbon carriers after 1\,Myr in these low CR rate models. 

When no significant conversion of CO into other carriers occurs, the CO snowline location is predominantly determined by the thermal structure of the disk. 
DM Tau is the coldest disk in our sample and its CO mid-plane snowline is around 10-15 AU in the standard model. In the TW Hya model, the CO mid-plane snowline is around 20\,AU, consistent with 21$\pm$1.3\,AU derived from $^{13}$C$^{18}$O (3-2) line observations \citep{Zhang17} and the N$_2$H$^{+}$ line image \citep{Qi13}. For the HD 163296 model, its CO mid-plane snowline is between 50-70\,AU, consistent with the previous estimation from \ce~and N$_2$H$^{+}$ line observations, after correcting to the Gaia2 distance \citep{Qi15}.  For the IM Lup model, its CO ice starts to appear around 30\,AU, but its CO ice column density increases slowly with radius compared with other disks, because the IM Lup disk is highly flaring and has a thicker CO gas layer than others. 

In short, the four standard models ($\zeta_{\rm CR}=1.36\times10^{-18}$ s$^{-1}$) all show a layered CO gas structure and the mid-plane CO snowline locations are consistent with existing observational constraints.  In these models,  less than 10\% of total CO (gas+ice) have been converted to other carbon carriers after 1\,Myr. 

\subsection{ CO abundance structures in models with a high cosmic-ray rate}

\begin{figure*}
\centering
\includegraphics[width=\textwidth]{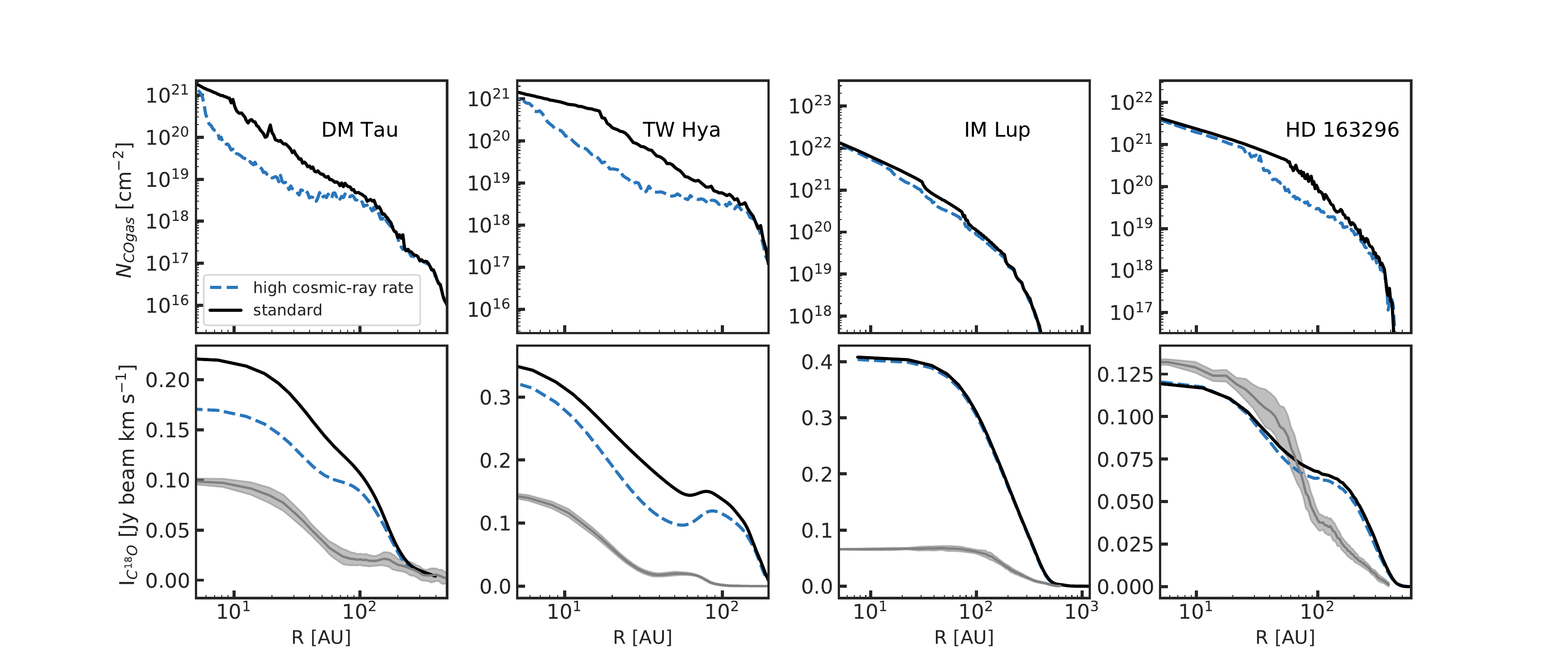}
\vspace{-0.1cm}
\caption{Comparison between high cosmic-ray ionization models and standard models ($\zeta_{\rm CR}=1.36\times10^{-17}$ s$^{-1}$ vs. $1.36\times10^{-18}$ s$^{-1}$). Top: CO gas column density in high $\zeta_{\rm CR}$ (blue dash line) and standard models (black solid line). Bottom: \ce~line surface brightness profiles in the models with different $\zeta_{\rm CR}$ rates (blue dash and black solids lines). The observed profiles are in grey.\label{fig:zeta}}
\end{figure*}

Figure~\ref{fig:zeta} shows the comparison of CO gas column density distribution in high and standard models ($\zeta_{\rm CR}=1.36\times10^{-17}$ s$^{-1}$ vs. $1.36\times10^{-18}$ s$^{-1}$).  Models of DM Tau and TW Hya show substantial CO gas depletions between 10-100\,AU, by a factor of 5-20 after 1\,Myr.  The HD 163296 model shows an intermediate level of depletion by up to a factor of 10 between 30-200\,AU. The IM Lup model shows the smallest variation between high and low CR rate models with a depletion factor less than 4 throughout the whole disk.  

In general, the chemical processing rates of CO gas depend on the thermal and ionization structures of disks which in turn depends on many physical properties, including the stellar luminosity, UV luminosity, disk flaring structure, and dust size distribution. For a given cosmic ray rate, the available surface areas on grain surface have a significant impact on the CO processing rate. \citet{Schwarz18, Bosman18} have carried out systematic studies on how disk properties would impact the CO depletion level, including small/large grain mass ratio and temperature. In our models, there are some very noticeable effects. The effect of grain surface area is most obvious in the IM Lup disk models that have only 1\% of the solid mass in small grains. Limited by the available grain surface areas, the chemical processing of CO in the IM Lup models is order(s) of magnitude smaller than that of other disks. The effect of thermal structure can be seen in HD 163296 models. As a Herbig disk, its gas temperature is much warmer than T Tauri disks and thus has less CO freeze-out regions for CO to be processed in the grain-surface path. 

Despite significant decreases in the CO gas column density, Figure~\ref{fig:zeta} shows that the surface brightness profiles of \ce~line emission are not very sensitive to the CO depletion caused by high CR ionization. This is because most of the depletion occurs in the deep region of these disks and the CO emission above the depletion region is already optically thick. These results indicate that the CO depletion by chemical processing alone, even by high CR rate, is unlikely to account of the low surface brightness observed in these four disks if their gas-to-dust ratio is still close to 100.

\subsection{ How the CO surface brightness profile changes with gas mass distribution}

\begin{figure*}[!htbp] 
\centering
\includegraphics[width=\textwidth]{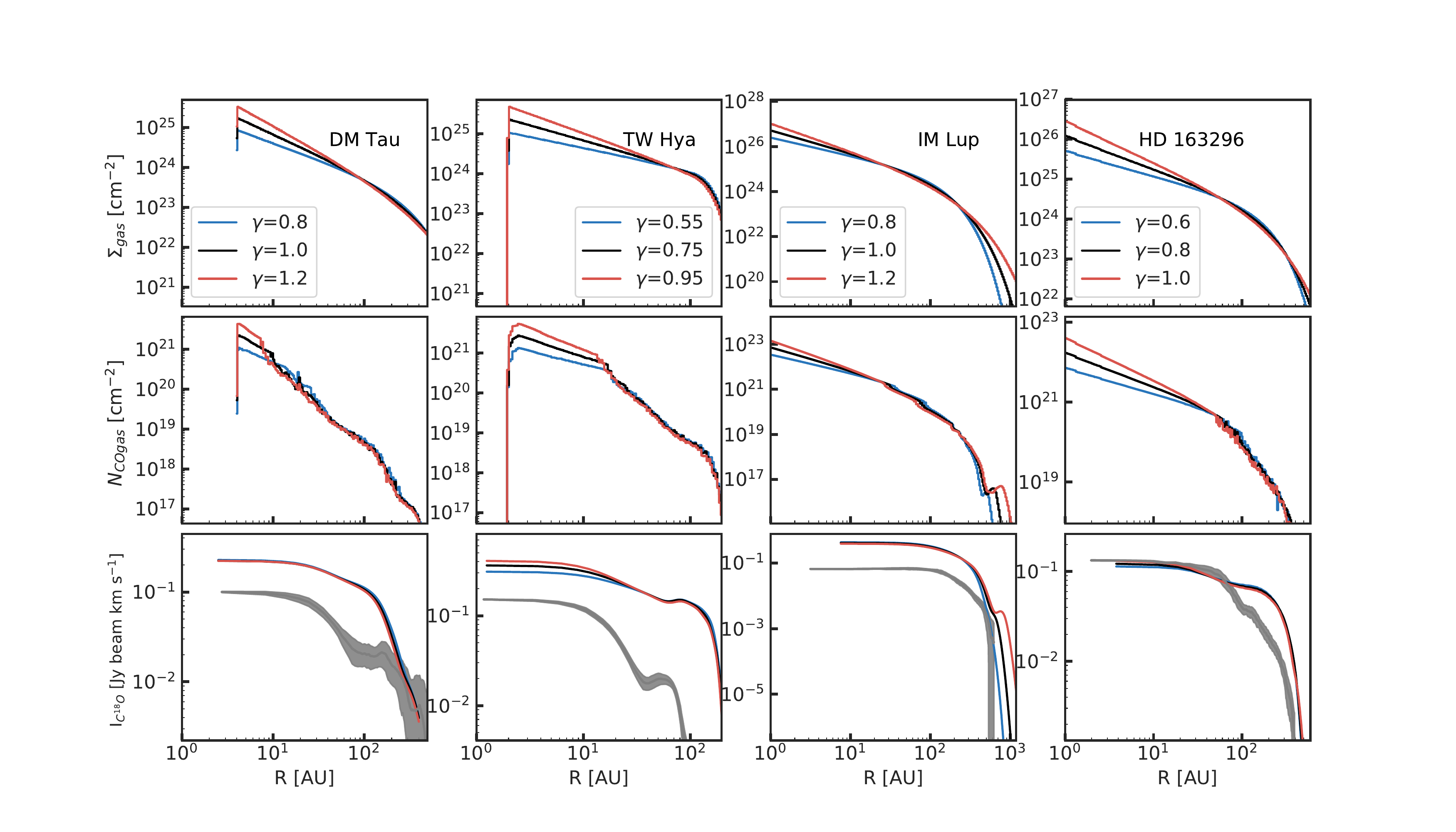}
\vspace{-0.5cm}
\caption{Top row: The total gas surface density profiles for models with different $\gamma$. Middle row: the CO gas column density profiles after 1\,Myr chemical evolution and a $\zeta_{\rm CR}$ of 1.36$\times10^{-18}$ s$^{-1}$.  Bottom row: model \ce~line brightness distributions vs. observations (grey line). \label{fig:gamma}}
\end{figure*}

Figure~\ref{fig:gamma} shows how the profiles of total gas surface column density ($\Sigma_{\rm g}$), CO gas column density ($\Sigma_{\rm CO}$), and the surface brightness of CO isotopologue line emission  ($I_{\rm CO}$) vary in models deviating from the $\gamma$ of standard models by $\pm$ 0.2. These results show that $\Sigma_{\rm CO}$ represents well the gas column density change within the mid-plane CO snowline, but it becomes less sensitive to the gas surface density changes beyond the CO snowline as the CO gas layer only accounts for a  small fraction of the total gas column. The $I_{\rm CO}$ profiles only change less than 30\% for uncertainty of 0.2 in the $\gamma$ index.

\subsection{Radial distribution of CO gas abundance: \\ models vs. observations}

\begin{figure*}[!htbp] 
\centering
\includegraphics[width=0.9\textwidth]{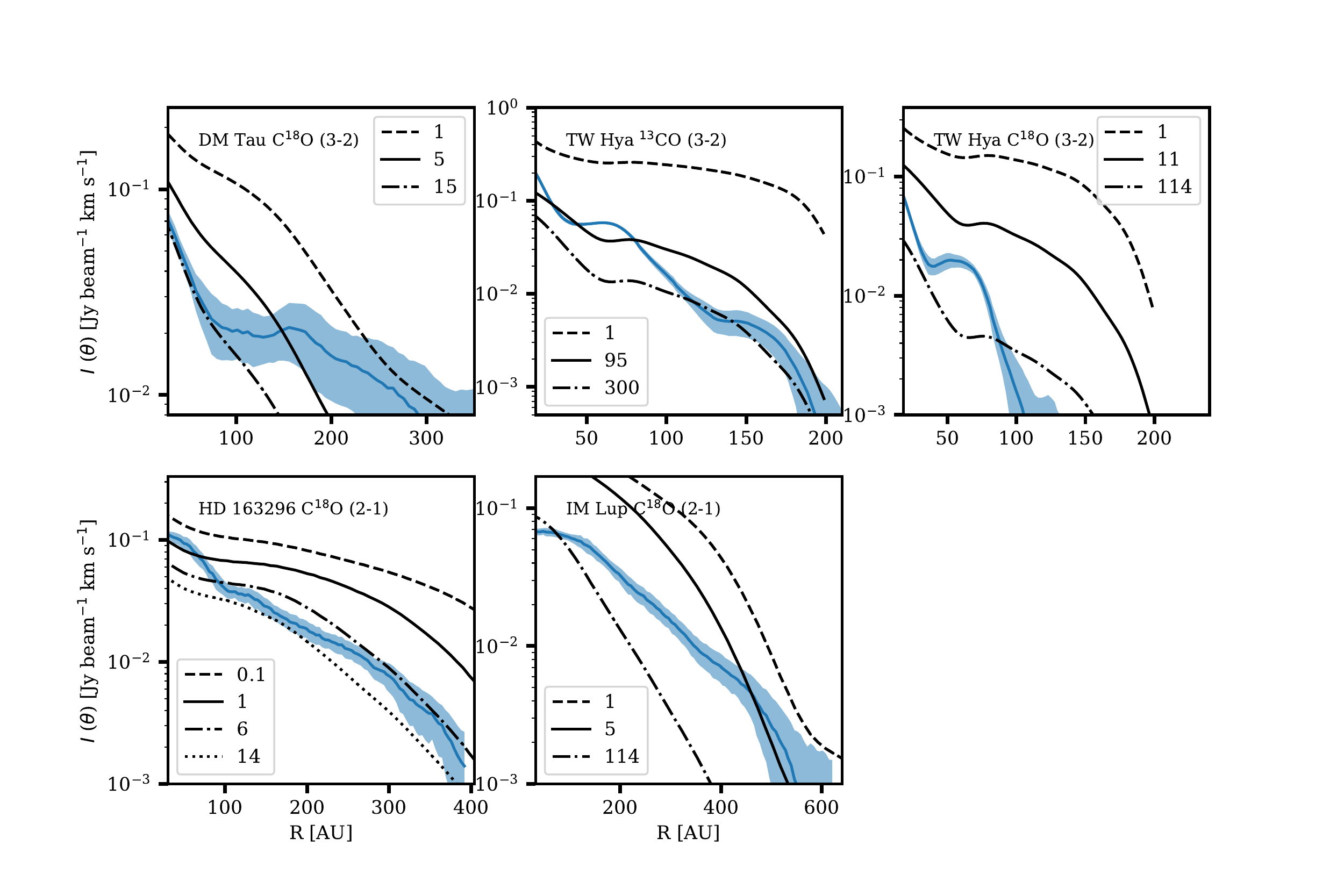}
\vspace{-0.cm}
\caption{Comparison of observed radial intensity profiles to models with different CO depletion factors. The observed profiles and their 1$\sigma$ uncertainties are in light blue and the models are in black. \label{fig:co_r_profile} }
\end{figure*}

Here we compare the surface brightness distributions of CO lines ($I_{\rm CO}$) between observations and our standard models to estimate the levels of variations in the radial CO gas abundance. For each disk, we generate a set of synthetic line images by rescaling the model CO gas abundance throughout the disk with a constant factor. The grid spans from a depletion factor of 0.1 to 400,  increasing by a factor of 1.2 at each step.  Figure~\ref{fig:co_r_profile}  shows the line surface brightness profiles of models with exemplary depletion scales, comparing with the observed profiles in the four disks. Using the grid of models, we record which depletion factor matches the best with models at each radius. 

\begin{figure}[!htbp] \centering
\includegraphics[width=0.5\textwidth]{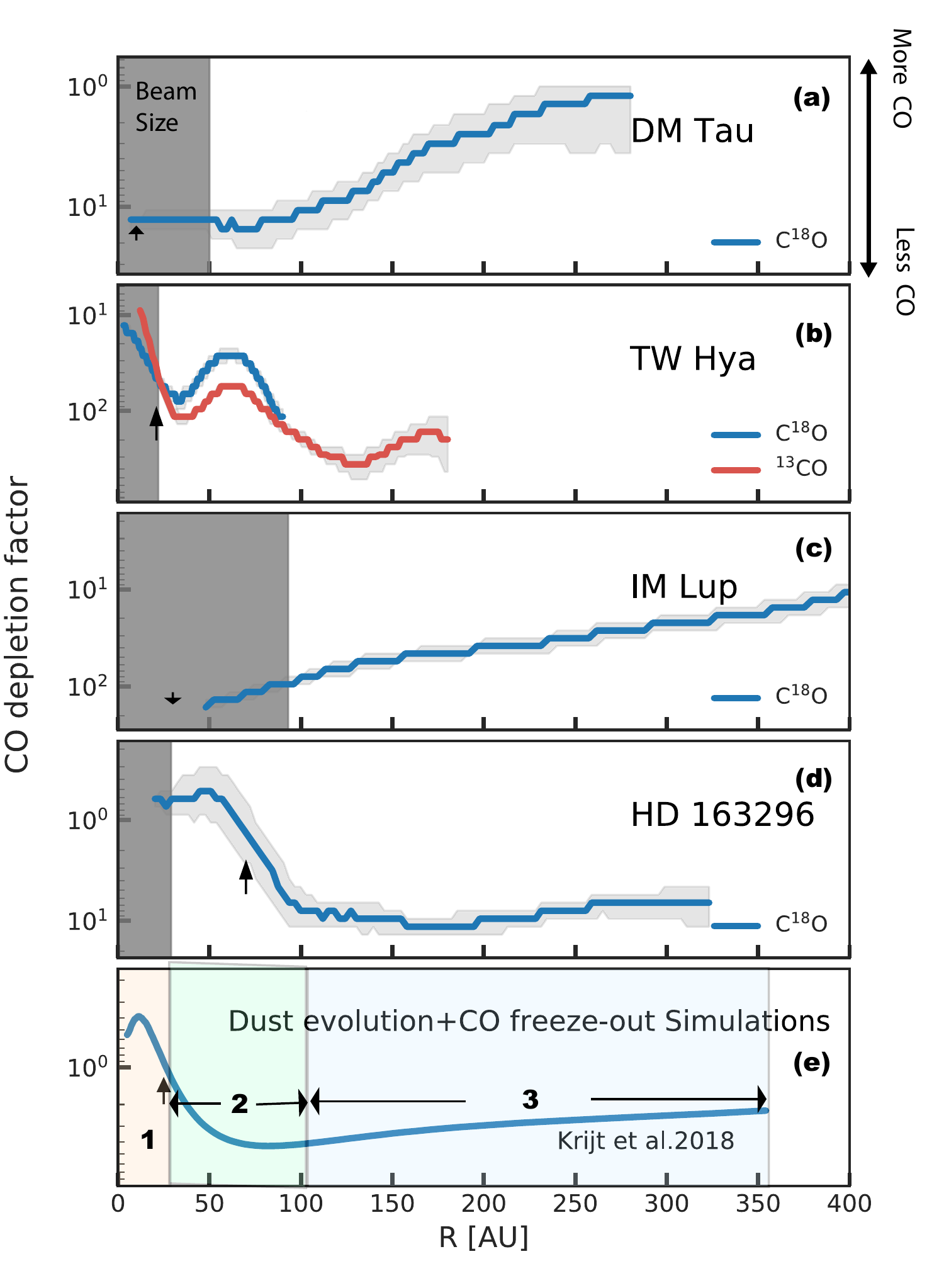}
\vspace{-0.7cm}
\caption{(a-d) The radial profiles of CO depletion factor in four disks. The black arrow indicates the location of CO snowline at the disk mid-plane and the grey shaded areas indicate the beam size of the observations. (e): the radial profile of CO depletion of the warm molecular layer in the 2-D dust evolution simulations of \citet{Krijt18}. The transparent colored boxes indicate three regions with different CO gas abundances: (1) the enhancement of CO gas abundance inside the mid-plane CO snowline, (2) the largest CO depletion region, (3) the CO gas abundance slowly increases with radius \label{fig:co_distri}. 
 }
\end{figure}

Figure~\ref{fig:co_distri} shows the required CO depletion distributions to match observations. 
On the disk-averaged scale, we find that significant depletion of CO gas abundance is needed to explain observations. 
This is consistent with the previous results of disk-averaged CO abundance studies \citep{Favre13, McClure16, Flaherty15, Cleeves16}. For the radial variation in individual disks, we find that one order of magnitude radial variation is usually needed to account for observations. Comparing to the changes of $I_{\rm CO}$ due to $\gamma$ uncertainty, it is unlikely these large deviations can be explained by surface density profiles alone.  

For individual disks, the DM Tau disk shows a nearly constant level of CO depletion inside 100\,AU (by a factor of 12) and its CO abundance gradually increases with radius to an interstellar ratio at 280\,AU.  TW Hya shows a more complicated pattern, with a sharp decrease of CO abundance between 10 and 30\,AU, and then increases with the radius until 60\,AU, and outside 100\,AU the CO abundance is extremely depleted ($>$ a factor of 200).  IM Lup shows a monotonic trend of CO abundance depletion, with an outer disk that has less depletion. HD 163296 shows a modest CO abundance enhancement inside 70\,AU (by a factor of 2-5),  its 100-200\,AU shows the largest depletion and then a shallow increase of CO abundance with radius. 

Except for the oldest disks TW Hya, all other three disks show a similar pattern in their CO abundance structures in regions beyond their CO snowline --- the CO gas abundance is  low in intermediate disk region outside the mid-plane CO snowlines, but gradually increases with radius to a higher value in the outermost region. Interestingly, this pattern is consistent with the general trends of carbon elemental abundance suggested by other carbon carriers.  \citet{Bergner19} analyzed C$_2$H radial brightness distributions in the same three disks and their models required a high depletion of elemental carbon abundance (by a factor of 10-100) between 100-200\,AU and only a modestly depletion beyond 200\,AU (by a factor of 1-10). Therefore, the CO depletion profiles likely represent a general depletion of carbon elemental abundance in the atmospheres of these disks.  

\textbf{Summary of model results:}  1. In detailed thermo-chemical models for the four disks, a high CR ionization rate can significantly reduce the CO gas column between 50-100\,AU up to a factor of 20 after 1\,Myr. However, the depletions by chemical processing mostly happen at the deep interior of these model disks, and the resulting CO abundance structures cannot reproduce the weak CO isotopologue lines observed. 2. Comparing these thermo-chemical models to spatially resolved observations, we find that the CO gas abundance likely varies significantly with radius. An interesting pattern in the radial profiles is that the CO gas abundance is  low in intermediate disk region outside the mid-plane CO snowlines, but gradually increases with radius to a higher value in the outermost region.

\section{Discussion}
\label{sec:discussion}

\subsection{CO depletion mechanisms}
Current surveys of nearby star formation regions suggest that gas mass derived from CO observations is 1-2 orders of magnitude too low compared to the canonic CO ratio in ISM \citep{Ansdell16, Long17, Miotello17}.
 In a broad sense, two types of depletion mechanisms have been proposed: (1) the first one is chemical processes that convert CO into other less volatile molecules (e.g., CO$_2$, CH$_3$OH, CH$_4$) \citep{Aikawa99_abundance, Bergin14, Reboussin15, Eistrup16, Yu16, Schwarz18, Bosman18};  (2) the second type is physical processes that CO gas abundance is altered as a result of the CO freeze-out onto dust grains and the subsequent dust growth, settling, and radial drifting in protoplanetary disks \citep{Kama16, Xu17, Krijt18}.  In other words, CO becomes trapped as ice in large grains that reside in the midplane or perhaps even in planetesimals. Previous studies of CO depletion mechanisms were mostly theoretical works. Here we discuss our results of radial CO variations in the context of these mechanisms proposed. 

\subsubsection{Chemical depletion mechanisms}
\label{sec:co_chemi}
Several chemical pathways have been proposed to account for CO depletion in protoplanetary disks. In gas-phase reactions, CO can be destroyed by He$^+$ and subsequently form CH$_4$ and hydrocarbons \citep{Aikawa99_abundance, Eistrup16, Yu16, Schwarz18}.  In grain-surface reactions, CO ice can react with OH to form CO$_2$, or CO ice goes through a sequence of hydrogenation to form CH$_3$OH \citep{Yu16, Schwarz18, Schwarz19a, Bosman18}. 

All of these destruction pathways require dissociation or ionization radiation, either from UV, X-ray, or cosmic-ray.  In general, models suggested that a CR ionization rate larger than 10$^{-17}$ s$^{-1}$ is needed to make an order of magnitude depletion of CO gas over a few Myr timescale of the gas lifetime in disks \citep{Schwarz18, Bosman18}. However, it is unclear that a high CR rate is present in actively accreting T Tauri disks,  because of cosmic-ray exclusion by winds and/or strong magnetic fields \citep{Cleeves13a, Cleeves15}. 

Previous chemical studies focused on the total gas column density of CO but did not consider how the vertical location and optical depth would impact the observed CO line intensities.  As we showed in Figure~\ref{fig:zeta}, although the CO gas column density decreases by a factor of 10 between 50-100\,AU in the high CR rate models of the TW Hya and DM Tau disks, the surface brightness of their \ce~line emission only changes less than 20\%, because the CO processing preferentially occurs at the deep layer while the CO gas abundance near the surface layer is not significant depleted.

Therefore, for relatively massive disks ($\ge$0.01\,M$_{\odot}$),  even when an ISM level CR rate is present and causes an order of magnitude depletion in the total CO gas column density, the predicted \ce~lines are still too strong compared with observations. 

We will discuss ways to increase CO depletion at the surface layer in section~\ref{subsec:enhanced_depletion}. 

\subsubsection{Physical depletion mechanisms}
Alternatively, the CO gas depletion may be a result of dust evolution in disks. Detailed models of dust evolution --- including grain coagulation, fragmentation, and settling combined with viscous gas mixing to deeper layers where grains are growing --- have shown the settling and radial drifting of icy pebbles can change the vertical and radial distribution of the CO gas in the disk. Indeed, 1D-models showed that the vertical mixing and dust settling at a single radius can deplete the gas-phase H$_2$O and CO abundance in the warm molecular layer by a factor of few to 50,  depending on the strength of turbulence and the location of the surface CO snowline \citep{Krijt16, Xu17}.   

\citet{Krijt18} constructed a 2D model that includes both vertical settling and radial drifting processes to investigate how the CO gas abundance evolves across the disk. In the vertical direction, they confirmed that within 1\,Myr, dust evolution can remove up to 80\% of the original CO vapor in the warm molecular layer outside the CO mid-plane snowline. In the radial direction, they showed that pebble formation and motion in the disk makes an interesting pattern on the radial distribution of CO gas abundance in the warm molecular layer (see Figure~\ref{fig:co_distri}): (1) in regions inside  the mid-plane CO snowline, CO gas abundance is enhanced by a factor of few compared to its initial value, (2) the largest CO gas depletion occurs at an intermediate region beyond the mid-plane CO snowline and peaks at a radius around a factor of 2-3 larger than the CO snowline, (3) in the outermost region beyond the CO depletion peak, the CO gas abundance smoothly increases with radius.

This radial pattern emerges because the CO depletion level at a given radius depends on the local timescales of dust growth and vertical mixing, and the supply of CO from other radial zones. The dust growth and vertical mixing timescales are expected to be proportional to the local orbital timescale \citep{birnstiel12, Krijt16}. The intermediate region beyond the CO mid-plane snowline has shorter timescales compared with that of the outermost region, and therefore its CO gas depletes faster than that of the outermost region.  The region just beyond the mid-plane CO snowline does not have the highest depletion because a fraction of CO vapor inside the snowline can diffuse into this region. The enhancement of CO gas at the inner disk region occurs when sufficient pebbles with a CO-ice mantle drift inwards into the region inside the CO snowline and evaporate CO into the gas phase. 

Comparing to our results, the depletion patterns of three disks (except for TW Hya)  are qualitatively consistent with this radial pattern due to dust evolution:  relatively large CO depletion beyond the mid-plane CO snowline and the outermost disk region is less depleted than the intermediate region. It is interesting that the CO depletion pattern in the HD 163296 disk is strikingly similar to the depletion pattern predicted by \citet{Krijt18}. In the HD 163296 disk, CO gas abundance appears to be enhanced by a factor of 5 inside its mid-plane snowline. This kind of enhancement is expected if grains with CO ice radially drift through the CO snowline and travels a large distance before they lose all of their CO ice. However, we do not see any significant enhancement of CO gas abundance in the inner region of DM Tau and IM Lup. 

Although the general radial dependence of CO depletion is qualitatively similar to the predictions of dust evolution, the absolute magnitudes of depletion found here are much higher than these in dust evolution simulations. We find a factor of 10 or larger CO depletion in most regions of the four disks, while the simulations of \citet{Krijt18} predicted a relatively modest depletion (by a factor of 2-6) in the most region of the disk.

\subsubsection{Ways to increase CO gas depletion in disks \label{subsec:enhanced_depletion}} 
As discussed above,  neither the chemical processing nor the dust growth alone can produce sufficiently large CO depletion at the observable layer to account for the weak CO emissions seen in Class II disks. 

There are several ways to increase the CO gas depletion in a disk. The simplest way is a longer dust evolution. The simulations in \citealt{Krijt18} stopped at 1\,Myr, and the CO depletion gradually increases with time. However, the depletion will eventually reach a steady state when most of the solids become pebbles and stay in the mid-plane region. The maximum depletion level depends on various disk properties such as the turbulence level and total dust masses. 

A more promising way is that there is enough vertical mixing between the atmosphere and the active CO processing region at the deep layer. The efficiency of this way can be easily seen by comparing the timescales of vertical mixing and chemical processing. 
The turbulent mixing timescale is given by $t_{\rm mix}\simeq H^2/D\sim(\alpha \Omega_k)^{-1}$, where $\alpha$ is the Shakura-Sunyeav $\alpha$ parameter and $\Omega_k$ is the Keplerian angular velocity \citep{Xu17, Krijt18}. For a disk around a solar mass star, $t_{\rm mix}\simeq10^5$\,yr at 100\,AU for $\alpha=10^{-3}$, and 10$^6$\,yr for $\alpha=10^{-4}$. The mixing timescales become only 10$^{4}$ and 6$\times$10$^{4}$\,yr for $\alpha=10^{-3}$ at 20, 50\,AU, respectively. Compared to chemical timescales of CO processing, \citet{Bosman18} found that the most efficient CO depletion occurs at regions between 15-30\,K with a typical timescale of 10$^{6}$\,yr. Therefore the mixing timescale is in general much shorter or comparable to the chemical timescale of CO depletion.  A coupling of mixing and chemical processes can significantly reduce the CO gas abundance at the observable layers, especially for regions inside 50\,AU.  Therefore, we suggest that a coupled view of physical-chemical evolution is necessary for future investigation of the CO depletion problem and even general understanding of volatile abundances in planet-forming regions.

\subsection {Effects of substructures on CO gas abundance profiles\label{dis:substructure_effects}}

Recent ALMA long-baseline observations have revealed that many disks show 1-20\,au scale substructures in their dust continuum emission \citep{Andrews18b}. Prominent substructures in the gas and dust mass distributions may alter the local CO gas emission. For example,  \citet{Facchini18} found that the gas inside gaps opened by 1-5\,M$_J$ planet becomes colder than the surrounding environment in their thermal-chemical models. The depletion of gas and temperature variation across gaps can affect a radial profile of the CO abundance derived from a smooth surface density structure such as ones used here. 

Although the origins of these dust substructures are still under debate, proposed mechanisms generally predict that substructures in CO line emission would be much narrower and shallower than these of the dust emission \citep{Facchini18}.
Three of the four disks (TW Hya, DM Tau, and IM Lup) only show modest substructures in continuum emission beyond 20 au, with up a factor of 3 depletion inside gaps \citep{Huang18b, Kudo18}. HD 163296 has the most significant substructures among the four disks, showing a factor of 33 depletion inside one of its gaps in the 1.3\,mm continuum \citep{Huang18b}.  But even for the HD 163296 disk, its \cc~and \ce~(2-1) line observations at 0.\arcsec2 resolution only show subtle breaks in the slopes rather than clear gaps \citep{Isella16}. In short, based on the depths and widths of known dust substructures in the four disks and current theoretical predictions, it is unlikely that the order of magnitude variations in CO abundance profiles are due to substructures alone.

On the other hand, it is  important to study how the dust substructures may affect the local chemical substructures and as these local effects may significantly impact the compositions of forming-planets. A Large ALMA Cycle 6 program is currently ongoing to study the detailed interplay between dust substructures and local chemical variations. 

\subsection{ The distribution of CO gas abundance and disk properties}

In our small sample, the radial profiles of CO depletion are diverse. Here we discuss how disk properties might play a role in regulating the radial CO depletion profile. 

HD 163296 is the only Herbig disk in our sample and it shows a very similar depletion pattern to the dust evolution models of \citet{Krijt18}. Assuming a canonical gas-to-dust mass ratio of 100, it has a modest CO gas depletion in this sample (by a factor of 5 on disk-average).  Current chemical models suggest that the chemical processing of CO gas is most efficiently happen at the cold interior of a disk. Herbig disks have a larger fraction of the disk mass above the CO condensation temperature than that of T Tauri disks,  and therefore they are expected to have less CO gas processing through the chemical pathways \citep{Bosman18}. Besides HD 163296, another example of Herbig disk with modest carbon depletion is HD 100546. \citet{Kama16} analyzed major carbon carriers (CO, CI, CII) for the HD 100546 disk, and concluded that the elemental carbon abundance in its warm molecular layer is close to the interstellar ratio or only modestly depleted, with C/H = 0.1-1.5 $\times$10$^{-4}$.  Considering the old ages of the two Herbig disks (HD 163296 $\simeq$12\,Myr, HD 100546 between 4.1-16\,Myr, \citealt{Andrews18b,vanderMarel19}), the chemical and physical processes should already have sufficient time to occur. The fact that the CO depletion is modest in these two warm disks suggests that temperature may play an important role in CO depletion processes.  A larger sample of Herbig disks can easily confirm or rule out the effect of warm temperature in the CO depletion observed in protoplanetary disks. 

Among the three T Tauri disks studied here, IM Lup is the most massive disk. Although it appears to be relatively young ($\sim$1\,Myr), its disk-averaged CO depletion is high, by a factor of 20 \citep{Cleeves16}.  Our model predicts that even with high cosmic-ray ionization rate, the expected CO depletion is a factor of 2-4. On the other hand, the high CO depletion in IM Lup is consistent with the expectation that dust evolution is faster in a massive disk. 

DM Tau is the coldest disk in our sample but its disk-averaged CO depletion is just modest. This appears to be inconsistent with that chemical processing of CO happens most efficiently at the cold region of a disk. On the other hand, the DM Tau disk is among the largest disks in both gas lines and (sub)mm continuum observations, and its column density is generally lower than those of other three disks (see Figure~\ref{fig:gamma}). The low-density environment can lead to slower dust growth and less CO depletion. 

The TW Hya disk shows the largest CO gas depletion on disk-average. In particular, its outer disk region ($>$100\,AU) appears to be extremely depleted of CO (by a factor of 100-300). This extreme CO depletion in the outermost region is puzzling, as it can be explained by neither chemical processing nor dust evolution. In particular, molecular ion observations towards the TW Hya disks indicate its ionization level due to CR is quite low ($\zeta_{\rm CR}<10^{-19}$ s$^{-1}$) \citep{Cleeves15}.  On the dust evolution side, the highest depletion occurs in the outermost region, which is inconsistent with the predictions of dust evolution simulations. 

Given these arguments, we suggest that the weak CO emission beyond 100\,AU of the TW Hya disk is more likely to be a result of gas dissipation beyond 100\,AU.  TW Hya is one of the oldest T Tauri disks (5-10\,Myr old) and thus is likely more evolved than other systems. Although its HD (1-0) line flux suggests its total gas disk mass is still high, the HD line mostly traces the relatively warm region inside 50\,AU \citep{Bergin13, trapman17}.  A lower limit of the gas mass beyond 100\,AU in the TW Hya disk has been constrained by ratios of resolved CS lines \citep{Teague18b}.  Comparing the difference between the minimum mass and the one used in our model, the CO abundance can be as high as only modestly depleted by a factor of 2-3. In summary, a gas-poor disk beyond 100\,AU is consistent with current observations of the TW Hya disk.

\section{Summary\label{sec:summary}}

Using well-resolved CO isotopologue line images and detailed thermo-chemical models, we investigate whether the CO gas depletion varies with radius in four well-studied protoplanetary disks, and compare these CO depletion profiles with predictions of chemical processes and dust evolution in protoplanetary disks. In summary, we find: 

\begin{itemize}
\itemsep0em 

\item The CO gas abundance in the warm molecular layer likely varies significantly with radius in these four disks. We find at least one order of magnitude variation within individual disks. 
 
 \item Our models show that chemical processing of CO without vertical mixing even with an ISM level cosmic-ray ionization rate, cannot explain the weak CO isotopologue lines observed in these disks. The reason is that the ionization driven CO depletions are active within the deep denser regions close to the mid-plane of these disks. Without sufficient vertical mixing, the chemical processing cannot cause orders of magnitude weaker CO isotopologue line emissions. 
    
 \item The general trend of the radial profiles of CO depletion in three disks (except for TW Hya) is qualitatively consistent with predictions of dust evolution models:  the outermost disk region generally has higher CO gas abundance than the intermediate region beyond the mid-plane CO snowline.  But the observed CO depletions are a factor of few higher than these in numerical simulations of dust evolution. 

\item In the HD 163296 disk, we find its CO gas abundance is slightly enhanced inside its CO snowline at 70\,AU. Unlike other three disks which generally have less depleted CO in their outermost region, the TW Hya disk shows a distinctive case. Its 100-200\,AU region is either extremely CO depleted or already lost a significant amount of gas. 

 \item We find that neither the dust growth nor chemical processing alone can reproduce the CO depletion levels seen in these four disks. We suggest that a coupled view of physical and chemical evolution is necessary to investigate the CO depletion problem and in general to understand what raw materials are available for planet formation at different distances from the central star.
\end{itemize}

\acknowledgments This paper makes use of the following ALMA data:
\dataset[ADS/JAO.ALMA\#2013.1.00226.S]{https://almascience.nrao.edu/aq/?project\_code=2013.1.00226.S},\\ 
\dataset[ADS/JAO.ALMA\#2013.1.00601.S]{https://almascience.nrao.edu/aq/?project\_code=2013.1.00601.S},\\ 
\dataset[ADS/JAO.ALMA\#2013.1.00798.S]{https://almascience.nrao.edu/aq/?project\_code=2013.1.00798.S},\\ 
\dataset[ADS/JAO.ALMA\#2015.1.00308.S]{https://almascience.nrao.edu/aq/?project\_code=2015.1.00308.S},\\ 
\dataset[ADS/JAO.ALMA\#2016.1.01495.S]{https://almascience.nrao.edu/aq/?project\_code=2016.1.01495.S}. \\ 
We thank the referee for comments improving this paper. 
ALMA is a partnership of European Southern Observatory (ESO) (representing its member states), National Science Foundation (USA), and National Institutes of Natural Sciences (Japan), together with National Research Council (Canada), National Science Council and Academia Sinica Institute of Astronomy and Astrophysics (Taiwan), and Korea Astronomy and Space Science Institute (Korea), in cooperation with Chile. The Joint ALMA Observatory is operated by ESO, AUI/NRAO, and NAOJ. The National Radio Astronomy Observatory is a facility of the National Science Foundation operated under cooperative agreement by Associated Universities, Inc.  K.Z., S.K., K.S. acknowledges the support of NASA through Hubble Fellowship grant HST-HF2-51401.001, HST-HF2-51394.001, and HST-HF2-51419.001 awarded by the Space Telescope Science Institute, which is operated by the Association of Universities for Research in Astronomy, Inc., for NASA, under contract NAS5-26555. 
\vspace{5mm}
\facilities{ALMA}
\software{\texttt{CASA} \citep{McMullin07}, \texttt{RADMC3D} \citep{radmc3d}, \texttt{RAC2D} \citep{Du14}}

\bibliographystyle{aasjournal}

\end{document}